\documentclass{aa}
\usepackage{psfig}
\usepackage{epsfig}
\def \sax {BeppoSAX}
\def \degmark{^\circ}

\def \hcm {\hbox {\ifmmode $ atom cm$^{-2}\else atom cm$^{-2}$\fi}}
\def \arcmin {\hbox{$^\prime$}}
\def \arcsec {\hbox{$^{\prime\prime}$}}

\def\approxgt{\mathrel{\hbox{\rlap{\lower.55ex \hbox {$\sim$}}
        \kern-.3em \raise.4ex \hbox{$>$}}}}
\def\approxlt{\mathrel{\hbox{\rlap{\lower.55ex \hbox {$\sim$}}
        \kern-.3em \raise.4ex \hbox{$<$}}}}
\newcommand{\mc}{\multicolumn}
\usepackage{multirow}
\begin{document}

\thesaurus{
(12.04.2; 13.25.3)}

\title{The \sax ~ 1-8 keV cosmic background spectrum}

\author{A. Vecchi\inst{1}
         \and S. Molendi\inst{1}
         \and M. Guainazzi\inst{2}
         \and F. Fiore\inst{3}
         \and A.N. Parmar\inst{2}
}
\offprints{S. Molendi (silvano@ifctr.mi.cnr.it)}

\institute{
        Istituto di Fisica Cosmica ``G.Occhialini'', Via Bassini 15,
        I--20133 Milano, Italy
\and
        Astrophysics Division, Space Science Department of ESA, ESTEC,
              Postbus 299, 2200 AG Noordwijk, The Netherlands
\and
        BeppoSAX Science Data Center, Nuova Telespazio, via Corcolle 19
        I-00131 Roma, Italy
}


\maketitle

\markboth{The \sax ~ 1-8 keV background spectrum}
{The \sax ~ 1-8 keV background spectrum}

\begin{abstract}
  
  The spectrum of the 1.0--8.0~keV cosmic X-ray background (CXB) at 
  galactic latitudes $>$$\vert 25 \vert \degmark$ has been measured 
  using the \sax\ LECS and MECS instruments. 
  The spectrum is consistent with a power-law of photon index 
  $1.40 \pm 0.04$ and normalization 
  $11.7 \pm 0.5$~photon~cm$^{-2}$~s$^{-1}$~keV$^{-1}$~sr$^{-1}$ 
  at 1~keV. 
  Our results are in good agreement with previous 
  ROSAT PSPC (Georgantopuolos et al. 1996), ASCA GIS (\cite{miyaji98})
  and rocket (McCammon \& Sanders 1990) measurements.  
  On the contrary, previous measurements with the HEAO1 A2 
  (\cite{marshall80}) and the ASCA SIS (Gendreau et al. 1995)
  instruments, are characterized by normalizations which are 
  respectively 35\% and 25\% smaller than ours.

\end{abstract}


\section{Introduction}
The first measurement  of the cosmic X-ray background (CXB) dates back to the 
early 60s (Giacconi et al. 1962). This serendipitous discovery posed the
still unsolved problem of the origin of the CXB.
Later observations (e.g. Schwartz \& Gursky 1974) have shown that the CXB 
above 1 keV is highly isotropic. This, as well as other evidence, has led to 
the current understanding that the CXB, at energies above 1 keV,
is of extragalactic origin. 
High angular resolution X-ray observations, and the absence of  
distorsions in the cosmic microwave background spectrum 
(Fixsen et al. 1996), support 
the idea that the CXB above 1 keV is dominated by the integrated 
emission from faint sources, with a dominant contribution coming 
from Active Galactic Nuclei (AGN). 
Recent results from X-ray surveys show that the fraction of the 
CXB due to discrete sourcers is 70-80\% in the 0.5--2.0 keV band 
(\cite{hasinger98}) and at least 30\% in the 2--10 keV band     
(\cite{fiore99}). The CXB spectrum in the 3--50 keV band,
measured by  HEAO1 A2 (\cite{marshall80}), is adequately represented by 
a power-law with an exponential cutoff at $\sim$ 40 keV. In the 2--10 keV  
energy range the CXB spectrum is approximated by a power-law with photon
index 1.4.
Further measurements of the CXB spectrum, performed with the 
ROSAT PSPC (\cite{hasinger92}) in the 0.5--2.0 keV band, with the ASCA
SIS (\cite{gendreau95}) in the 0.4--7.0 keV band, and with the 
ASCA GIS (\cite{miyaji98}) in the 1.0--10.0 keV band,
disagree at about the 20\%-30\% level as to the value of the 
1 keV normalization.
While the ASCA GIS and ROSAT PSPC measurements
give values around 11~photon~s$^{-1}$~cm$^{-2}$~keV$^{-1}$~sr$^{-1}$,
the ASCA SIS spectrum, which lies on the extrapolation of the  HEAO1 A2
measurement, yields a value of 
$\sim$ 9~photon~s$^{-1}$~cm$^{-2}$~keV$^{-1}$~sr$^{-1}$. New 
observations of the CXB spectrum are needed to clarify the issue.

In this letter we present a new measurement of the CXB spectrum,
in the 1.0-8.0 keV, obtained with the LECS and MECS instruments on 
board \sax.
A previous measurement, using LECS data alone, has been presented 
by Parmar et al. (1999; P99 hereinafter).
The remainder of the letter is organized as follows.
In Sect. 2 we discuss instrumental issues relevant to the analysis 
of the LECS and MECS CXB spectra. In Sect. 3 we present the 
list of observations which have been used to measure the  CXB spectrum.
In Sect. 4 we report on the spectral analysis. In Sect. 5
we discuss our findings and compare them to previous results. 
   
\section{Instrumental issues}
\label{sect:obs}

\subsection{LECS}

The Low-Energy Concentrator Spectrometer (LECS; 0.1--10~keV; 
Parmar et al. 1997) is an imaging scintillation proportional 
counter on-board the Italian-Dutch \sax\ X-ray astronomy mission 
(\cite{boella97a}). 
It has a circular field of view of 18.5\arcmin~ radius, an effective 
area of $\simeq$40~cm$^{2}$ at 2~keV and an energy resolution 
of $\simeq$8\% at 6~keV. Details on the CXB data reduction can be 
found in P99. Briefly, the CXB was accumulated 
within a central circle with an 8\arcmin ~ radius.
The spectrum of the non X-ray background (NXB) 
has been accumulated during 816.3~ks of dark Earth pointing 
(this is a factor 1.6 greater than in P99). 
The response matrix we employ is the same used by P99. 
Briefly, we recall that the response matrix, which is  
appropriate for the diffuse emission, 
has been generated, by simulating a set of 100 point sources, randomly
distributed within a radius of 12\arcmin, therefore larger than the CXB 
accumulation region. 
This matrix includes also the effect of off-axis mirror vignetting 
and the average obscuration of the support strongback.
Moreover the LECS matrix has been corrected
for the $\sim$15\% LECS/MECS crosscalibration missmatch present in the
September 1997 release of \sax ~ matrices. We expect any residual
crosscalibration error between LECS and MECS to be less than $\sim 5\%$.
In addition to the above effects, 
single reflected X-rays from within 120\arcmin\ can
be detected in the FOV (\cite{conti94}). 
The magnitude of this effect has been found to be $<$1\%
of the flux within an 8\arcmin\ extraction radius (P99).
Since this is well
within the uncertainty in CXB normalization, this effect is ignored.

\subsection{MECS}

The Medium-Energy Concentrator Spectrometer (MECS; 1.5--10~keV;
\cite{boella97b})
is an imaging gas scintillation proportional counter.
The MECS was originally composed of 3 units, MECS1, MECS2 and MECS3.
MECS1 failed in May 1997, in this paper we shall use only data from
the 2 units which are still operative. The  energy  resolution  
is 8\% at 6 keV.  
The combined on-axis effective area for the MECS2 and MECS3 units
is $\sim$80~cm$^2$ at 2.0 keV and $\sim$60~cm$^2$ at 8~keV. 
The MECS has a circular field of view (FOV) of 25$'$ radius.
An annular support structure, commonly referred to as strongback,
is localized at about 10$'$ from the center of the detector.  
The absolute flux calibration of the MECS was performed 
using the Crab nebula spectrum. Assuming a  power-law model 
the photon index, $\alpha$, and the 2-10 keV flux, F(2-10), 
for the Crab, are found to be  $\alpha = 2.088\pm 0.002 $  
and F(2-10)$=2.008\pm 0.006 \times 10^{-8}$ erg cm$^{-2}$s$^{-1}$. 
Repeated observations have not revealed any significant variations in 
either of these parameters so far (Sacco 1999).
The uncertainty in the line of sight N$_{\rm H}$ for the Crab,
may affect at the few percent level the MECS calibration 
in the softest (1.65-2.0 keV) energy band.
For the present work we have chosen to accumulate spectra from 
a relatively small central circular region with radius 8$'$. 
The main reasons  for this choice are that: 
{\it i)} the ratio of the CXB 
to the NXB is at its highest at the
center of the detector;
{\it ii)} we avoid complications associated to the obscuration 
form the strongback; 
{\it iii)} the inner region of the MECS detector is better calibrated 
than the outer one. 
We have created a response matrix appropriate for uniform diffuse emission. 
This matrix differs from the standard
point-source response matrix in the following ways:  
{\it i)}the effects of off-axis mirror vignetting are included;
{\it ii)}the CXB is modeled as a uniform diffuse emission,
  within a radius of 12\arcmin. 
The MECS response matrix has been generated using the {\sc EFFAREA}
program publically available within the latest {\sc SAXDAS} release.
The NXB accounts for a significant fraction of the total background,
thus a correct measurement of the NXB is needed to measure the 
CXB.
A total of 1850~ks of NXB data was accumulated using MECS dark Earth
pointings.
In the central 8\arcmin\ the MECS NXB spectrum is
approximately constant with energy, with a count rate of
$4\times 10^{-4}$~cts~s$^{-1}$~keV~$^{-1}$ MECS$^{-1}$, between 1.0 
and 4.5 keV. Between 4.5 keV  and 7 keV there is a
smooth increase in count rate with 3 discrete line-like features
superposed. 
Above 7 keV the spectrum is again flat with a count rate of 
$8\times 10^{-4}$~cts~s$^{-1}$~keV~$^{-1}$ MECS$^{-1}$. 
Above $\sim$8~keV, the NXB dominates the overall background spectrum.
Due to the
low-inclination, almost circular, \sax\ orbit,
variations in the NXB counting rate around the orbit are negligible.

\section {CXB observations}
\begin{table*}
\caption[]{\sax ~ observations used to create the
CXB spectrum. ${\rm N_{gal}}$ is the line of sight absorption in units 
of $10^{20}$~\hcm\ (\cite{dickey90})}
\begin{flushleft}
\begin{center}
\begin{tabular}{llrlllrrr}
\hline\noalign{\smallskip}
       \mc{2}{c}{Pointing (J2000)} 
       & ${\rm l_{II}}$ \hfil & \hfil ${\rm b_{II}}$ & \mc{2}{c}{Observations} 
       & ${\rm T^{LECS}_{exp}}$
       & ${\rm T^{MECS}_{exp}}$ &${\rm N_{gal}}$\\
       \hfil RA & \hfil Dec & $(\degmark)$ \hfil &
       \hfil $(\degmark)$ & \hfil Start  & \hfil Stop & (ks) & (ks) & \\
       \hfil (h~m~s) & \hfil ($\degmark\ 
       \arcmin\ \arcsec$) & & & (yr~mn~day) & (yr~mn~day) & & & \\
\noalign{\smallskip\hrule\smallskip}
01 39 35.4 & +89 14 06 & 123.1 & +26.4 & 1997 Feb 01 & 1997 Feb 03 & \dots & 
115.5 & 6.8 \\
02 31 42.0 & +89 15 47 & 123.4 & +26.5 & 1996 Jul 01 & 1997 May 23 & 69.7  & 
257.8 & 7.0 \\
05 52 07.9 & $-$61 05 35 & 270.0 & $-$30.6 & 1998 Mar 10 & 1998 Mar 22 & 46.7  
& 156.6 & 5.3 \\
06 12 22.7 & $-$60 59 04 & 270.1 & $-$28.2 & 1996 Oct 10 & 1996 Oct 12 & 43.3  
& 115.0 & 4.2 \\
06 22 36.4 & $-$69 15 28 & 279.5 & $-$27.8 & 1998 Oct 23 & 1998 Oct 23 & 12.4  
&  34.4 & 6.9 \\
16 35 10.7 & +59 46 30 &  89.9 & +40.2 & 1996 Aug 27 & 1996 Aug 31 & 79.9  & 
170.4 & 2.0 \\
16 51 19.5 & +60 11 49 &  89.8 & +38.1 & 1996 Aug 23 & 1996 Aug 27 & 82.3  & 
\dots & 2.1 \\
17 30 42.2 & +60 55 32 &  89.9 & +33.2 & 1996 Aug 21 & 1996 Aug 23 & \dots &  
65.8 & 3.5 \\
17 38 57.1 & +68 01 10 &  98.2 & +31.8 & 1999 Mar 03 & 1999 Mar 03 & 20.5  & 
\dots & 4.4 \\
17 49 33.7 & +61 05 54 &  90.0 & +30.9 & 1998 Mar 28 & 1998 Mar 29 & \dots &  
53.4 & 3.5 \\
17 50 51.1 & +61 05 45 &  90.0 & +30.8 & 1998 Sep 10 & 1999 Apr 11 & 88.1  &  
49.8 & 3.5 \\
17 52 07.4 & +61 01 01 &  89.9 & +30.6 & 1998 Aug 22 & 1998 Oct 07 &116.6  & 
206.0 & 3.5 \\
17 56 46.1 & +61 11 45 &  90.2 & +30.1 & 1997 Mar 23 & 1997 Mar 25 & 94.2  &
116.7 & 3.4 \\
17 58 14.6 & +61 12 30 &  90.2 & +29.9 & 1997 Mar 18 & 1997 Mar 20 & \dots & 
119.0 & 3.4 \\
18 18 20.5 & +60 58 42 &  90.2 & +27.4 & 1997 Apr 13 & 1997 Apr 15 & 26.6  & 
114.4 & 3.8 \\
23 07 53.5 & +08 50 06 &  84.4 & $-$46.1 & 1997 Dec 13 & 1997 Dec 14 & 13.9  & 
\dots & 4.7 \\
\noalign{\smallskip}
\hline
\end{tabular}
\end{center}
\end{flushleft}
\label{tab1}
\end{table*}


A background spectrum, with a total exposure time of 694.2~ks for the LECS
(a factor 1.25 greater than in P99)
and 1574.8~ks for the MECS, was accumulated from a number of
high galactic latitude exposures by extracting counts within an
8\arcmin\ radius of the nominal source position within the LECS and
MECS FOV (Table~\ref{tab1}).
While a light leakage problem has forced us to use the LECS 
only during night time, the MECS data has been accumulated
during day and night time.  
For various reasons of the 16 pointing listed 4 have only MECS 
data and 3 have only LECS data.
We note that the total integration times, the covered solid  
angles (0.67 sq deg for the LECS, and 0.73 sq deg for the MECS) 
and most importantly the accumulated photons (more than 44000 for 
the LECS plus MECS),  
are larger than those associated to 
previous measurements of the CXB spectrum, with imaging instruments,
in the medium energy band. For this reason we expect our measurement
to be limited by systematic effects rather than by  
counting statistics or by fluctuations associated to the 
CXB granularity.  
No point sources are present in the individual fields with 0.1--2.0~keV
fluxes $>$$1.7 \times 10^{-13}$~erg~cm$^{-2}$~s$^{-1}$,
and with 2.0--10.0~keV fluxes $>$$5 \times 10^{-13}$~erg~cm$^{-2}$~s$^{-1}$.
Using the 2-10 keV LogN-LogS reported in Comastri et al. (1999) 
we have estimated that about 5\% of the total CXB is produced
by source above the $5 \times 10^{-13}$~erg~cm$^{-2}$~s$^{-1}$
flux limit.  
All the fields have galactic latitudes $>$$\vert 25\degmark \vert$ 
and galactic column densities between 2.0 and 
$7.0\times 10^{20}$~\hcm. The spectrum was
rebinned to have at least 20 counts per bin 
in order to ensure the applicability of the $\chi ^2$ statistic. All
uncertainties are quoted at the 90\% confidence level for one
interesting parameter ($\Delta \chi^2 = 2.7$). 

\section{X-ray background spectrum}
\label{sect:bgfits}

Spectral fitting has been performed using the 1.0-4.0 keV LECS spectrum 
and the 1.65-8.0 keV MECS spectrum (as customary we have equalized MECS2 
and MECS3 data and produced a single MECS spectrum). Data below 1.0 keV is ignored 
as we are interested in characterizing the extragalactic CXB. LECS data 
above 4.0 keV, where a slight miscalibration problem is present, has 
also been ignored. 
MECS data above 8.0 keV, where the NXB becomes dominant is also
ignored.
All spectral models have been absorbed by a foreground column density,
N$_{\rm H}$, of $3.8\times 10^{20}$ atom cm$^{-2}$, which has
been derived by averaging the galactic densities of the 
blank fields listed in Table 1. 
In all fits we have allowed for a maximum cross calibration 
mismatch of 5\% between LECS and MECS data (see Sect. 2.1). 
A simple power-law fit gives an acceptable $\chi^2$ of 75.1
for 74 degrees of freedom (dof) with $\alpha = 1.40 \pm 0.04$, 
a normalization of
$11.7 \pm 0.5$~photon~s$^{-1}$~cm$^{-2}$~keV$^{-1}$~sr$^{-1}$ at
1~keV and a normalization of
$1.70 \pm 0.04$~photon~s$^{-1}$~cm$^{-2}$~keV$^{-1}$~sr$^{-1}$ at
4 keV. 
The normalization at 1 keV is quoted for comparison with
previous works, while the normalization at 4 keV is quoted because
when using a 1.0-8.0 keV spectral band the most precise determination
of the normalization is obtained at the center of the band rather than 
at one of its limits.

Since a fraction of the 1-2 keV CXB may quite possibly not be
of extragalactic origin, we have fitted the CXB spectra with a
power-law together with a single temperature optically thin 
plasma (the Mewe-Kaastra-Liehdal plasma emissivity model
in {\sc xspec}, \cite{mewe85}).
Since we are not interested in the thermal component in itself, 
we fixed its parameters to the best fitting values derived by  
P99 when fitting the LECS spectrum of the CXB. 
The  power-law plus thermal component fit, which is reported in Fig. 1,
gives an acceptable $\chi^2$ 
of 75.0 for 74 degrees of freedom (dof) with $\alpha = 1.35
\pm 0.04$, a normalization of
$11.0 \pm 0.5$~photon~s$^{-1}$~cm$^{-2}$~keV$^{-1}$~sr$^{-1}$ at
1~keV and a normalization of
$1.70 \pm 0.04$~photon~s$^{-1}$~cm$^{-2}$~keV$^{-1}$~sr$^{-1}$ at
4 keV.
Obviously, the addition of a low energy component has 
resulted in a reduction of the spectral index and of the 1 keV
normalization, while the 4 keV normalization is unaffected.

\begin{figure}
\centerline{
      \hbox{
      \psfig{figure=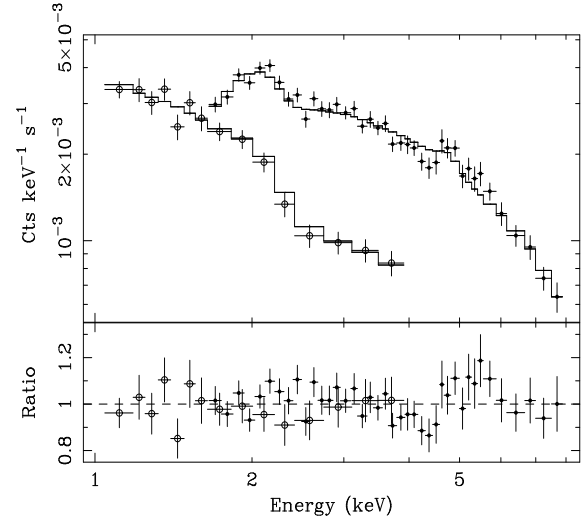,height=8cm,width=10cm,angle=-90}
}}
\caption{
Spectra and residuals in units
of the data/model ratio (lower panel) for the power-law
plus thermal emission model. Open circles are LECS data,
and filled circles are MECS data.
}
\label{figmg7}
\end{figure}

\section{Discussion}
In Fig. 2 we plot various measurements of the CXB spectrum.
To facilitate the comparison the spectra are plotted in the form 
of a ratio of the photon spectrum to a power-law with index 1.4
and normalization at 1 keV of 11~photon~s$^{-1}$~cm$^{-2}$~keV$^{-1}$~sr$^{-1}$.
Note that we have associated a 5\% error to the normalization of our 
measurement to account for residual uncertainties in the absolute calibration 
of the LECS and MECS instruments.

\begin{figure}
\centerline{
      \hbox{
      \psfig{figure=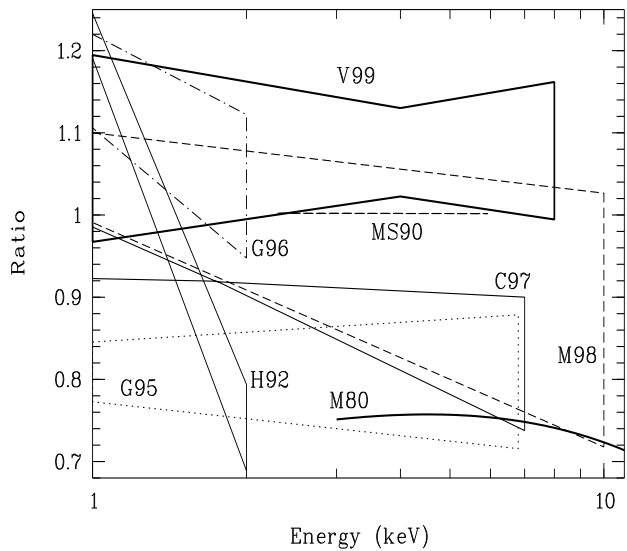,height=8.0cm,width=9.5cm,angle=-90}
}}
\caption{
Ratio of the CXB spectrum to a power-law with photon index 1.4
and normalization at 1 keV of 11~photon~s$^{-1}$~cm$^{-2}$~keV$^{-1}$~sr$^{-1}$. 
The thick solid line (M80) represents the best fit reported by Gruber 
et al. (1999) to the HEAO1 A2 measurement by Marshall et al. (1980). 
The thin solid horn (H92) is the ROSAT PSPC measurement
from Hasinger (1992).
The dot-dashed horn (G96) is a ROSAT PSPC measurement from 
Georgantopoulos et al. (1996).
The dashed line (MS90) is from rocket measurements 
(McCammon \& Sanders 1990).
The dotted horn (G95) is from the ASCA SIS 
measurement by Gendreau et al. (1995). 
The long-dashed horn (M98) is the ASCA GIS 
measurement by Miyaji et al. (1998) on the Lockman Hole. 
The  thin solid bowtie (C97) is a joint ROSAT
PSPC  ASCA SIS analysis of QSF3 by Chen et al. (1997) for E$>1$ keV.
Finally the thick solid bowtie (V99) is our 
own \sax ~LECS and MECS simple power-law fit
}
\label{fig3}
\end{figure}
The LECS measurement of P99 is consistent with the LECS-MECS measurement 
we present here, we have not included the P99 spectrum in Fig.2 to 
avoid making it even more crowded than it allready is.
Our measurement is consistent with the  ASCA GIS 
measurement of Miyaji et al. (1998) on the Lockman Hole. Their normalization 
at 4 keV is smaller than ours by $\sim$13\%, and within the  
errors. We are also in agreement with the rocket measurement
(McCammon \& Sanders 1990) and with the ROSAT PSPC measurement 
from Georgantopoulos et al. (1996).
The joint ROSAT
PSPC  ASCA SIS analysis of QSF3 by Chen et al. (1997) is
in agreement with our measurement at low energies but, due 
to its steeper spectral index, falls short of our measurement 
by about 25\% at 7 keV.
The ASCA SIS measurement of Gendreau et al. (1995) and the HEAO1 
measurement of Marshall et al. (1980) have spectral indices similar to 
ours, but normalizations which are smaller by $\sim$25\% and 
$\sim$30\% respectively. 
The discrepancy with the HEAO1 measurement is somewhat larger,
$\sim$35\% if we consider the contribution of bright sources
to the total CXB discussed in Section 3. 
    
\begin{acknowledgements}
\sax~ is a joint Italian-Dutch programme. MG acknowledges
snd ESA Fellowship.
We acknowledge support from the \sax ~ SDC.
We thank the referee, G.Hasinger for useful comments. SM
thanks A.S.Comastri for useful discussions.
\end{acknowledgements}


\begin{thebibliography}{}




\bibitem[Boella et al. 1997a]{boella97a}
Boella G., Butler R.C., Perola G.C., et al., 1997, A\&AS 122, 299

\bibitem[Boella et al. 1997b]{boella97b}
Boella G., Chiappetti, L., Conti, G., et al., 1997, A\&AS 122, 327


\bibitem[Chen et~al. 1997]{chen97}
Chen L.-W., Fabian A.C., Gendreau K.C., 1997, MNRAS 285, 449


\bibitem[]{}
Comastri A., Fiore F., Giommi P., et al., 1999, Ad.S.R., in press
(astro-ph/9902060)

\bibitem[Conti et al. 1994]{conti94}
Conti G., Mattaini E., Santambrogio E.B., et al., 1994,
SPIE 2279, 101


\bibitem[Dickey \& Lockman 1990]{dickey90}
Dickey J.M., Lockman F.J., 1990, ARA\&A 28, 215

\bibitem[Fiore et al. 1999]{fiore99}
Fiore F., La Franca F., Giommi P., et al., 1999, MNRAS, 306, L55 

\bibitem[]{}
Fixsen D.J., Cheng E.S., Gales J.M., et al., 1996, ApJ, 473, 576



\bibitem[Gendreau et al. 1995]{gendreau95}
Gendreau K., Mushotzky R., Fabian A.C., et al., 1995, PASJ 47, L5

\bibitem[Georgantopoulos et al. 1996]{georgantopoulos96}
Georgantopulos I., Stewart G., Shanks T., et al., 1996, MNRAS 280, 276

\bibitem[]{}
Giacconi R., Gursky H., Paolini F. \& Rossi, B., 1962, Phys. Rev. Lett.
9, 439

\bibitem[Griffiths \& Jordan (1998)]{jordan98} 
Griffiths N.W., Jordan C., 1998, ApJ 497, 883

\bibitem[]{}
Gruber D.E., Matteson J.L., Peterson L.E., Jung G.V. 1999, Report-no: SP-98-25
(astro-ph/9903492)

\bibitem[Hasinger 1992]{hasinger92}
Hasinger G., 1992, in: The X-ray Background, 
Barcons X., Fabian A.C., (eds.)
Cambridge University Press, Cambridge, p.~299



\bibitem[Hasinger et al. 1998]{hasinger98}
Hasinger G., Burg R., Giacconi R., et al., 1998, A\&A 329, 482








\bibitem[Marshall et al. 1980]{marshall80} 
Marshall F.E., Boldt E.A., Holt S.S., et al., 1980, ApJ 235, 4

\bibitem[McCammon \& Sanders 1990]{mccammon90}
McCammon D., Sanders W.T., 1990, ARAA 28, 657


\bibitem[Mewe et al. 1985]{mewe85}
Mewe R., Gronenschild E.H.B.M., van den Oord G.H.J., 1985, A\&AS 62, 197


\bibitem[Miyaji et al. 1998]{miyaji98}
Miyaji T., Ishisaki Y., Ogasaka Y., et al., 1998, A\&A 334, L13






\bibitem[Parmar et al. 1997]{parmar97} 
Parmar A.N., Martin D.D.E., Bavdaz M., et al., 1997, A\&AS 122, 309

\bibitem[Parmar et al. 1999]{parmar99} 
Parmar A.N., Guainazzi M., Oosterbroek T., et al., 1999, A\&A 345, 611
(P99)

\bibitem []{}
Sacco B., in BeppoSAX June 1999 EIWG meeting report 

\bibitem []{}
Schwartz D. \& Gursky H.  1974
in: X-ray Astronomy, Giacconi R., Gursky H.
(eds.) D. Reidel Publishing Company, Dordrecht, p.~359









\end{thebibliography}
\end{document}